\documentclass[11pt,preprintnumbers,superscriptaddress,nofootinbib]{revtex4}

\usepackage{url}
\usepackage{graphicx}
\usepackage{dcolumn}
\usepackage{bm}
\usepackage{amssymb}
\usepackage{amsmath}
\usepackage{epsfig}    
\usepackage{color}
\usepackage{slashed}
\usepackage{hyperref}
\usepackage{comment}
\usepackage[utf8]{inputenc}

\newcommand{\hcoup}{{\sc H-COUPv1} }
\newcommand{\GeV}{{\rm\ GeV}}
\newcommand{\sba}{s_{\beta-\alpha}}
\newcommand{\cba}{c_{\beta-\alpha}}
\newcommand{\tb}{\tan\beta}

\begin{document} 


\title{Loop effects on the Higgs decay widths in extended Higgs models}

\author{Shinya Kanemura}
\affiliation{Department of Physics, Osaka University, Toyonaka, Osaka 560-0043, Japan}

\author{Mariko Kikuchi}
\altaffiliation[Address after April 2018: ]{National Institute of Technology,
Kitakyushu College, 5-20-1 Shii, Kokuraminamiku,\\[-1.5mm] 
Kitakyushu, Fukuoka, 802-0985 Japan}
\affiliation{Department of Physics, National Taiwan University, Taipei 10617, Taiwan}

\author{Kentarou~Mawatari}
\affiliation{Department of Physics, Osaka University, Toyonaka, Osaka 560-0043, Japan}

\author{Kodai Sakurai}
\affiliation{Department of Physics, University of Toyama, 3190 Gofuku, Toyama 930-8555, Japan}
\affiliation{Department of Physics, Osaka University, Toyonaka, Osaka 560-0043, Japan}

\author{Kei Yagyu}
\altaffiliation[Address after April 2018: ]{Seikei University, Musashino, Tokyo 180-8633, Japan}
\affiliation{INFN, Sezione di Firenze, and Department of Physics and Astronomy, University of Florence, Via G. Sansone 1, 50019 Sesto Fiorentino, Italy}

\preprint{OU-HET 960}
\preprint{UT-HET 124}

\begin{abstract}
\begin{center}
{ \bf Abstract}
\end{center}
In order to identify the Higgs sector using future precision data, 
we calculate the partial decay widths of the discovered Higgs boson with the mass of 125~GeV
into fermion pairs and gauge-boson pairs
with one-loop electroweak and one-loop QCD corrections  
in various extended Higgs models,
such as the Higgs singlet model and four types of two Higgs doublet models.
In the tree-level analysis, 
the patterns of deviations from the standard model predictions in the partial decay widths for various decay modes are distinctive for each model, 
due to the mixing of the Higgs boson with other neutral scalars.
Our present analysis shows that even with a full set of radiative corrections 
we can discriminate these extended Higgs models via the partial decay widths 
as long as any of the deviations is detected at future precision measurements.
Furthermore, we quantitatively show that in each model 
the magnitude of the deviations can provide important information on the mass scale of extra Higgs bosons 
under the theoretical constraints from perturbative unitary and vacuum stability, 
which can be obtained without discovery of the additional Higgs bosons. 

\end{abstract}

\maketitle


\newpage
\section{Introduction}

Except a resonance with the 125~GeV mass, the LHC experiments have not observed any other states expected by new physics (NP) beyond the standard model (SM),
and instead put limits on them to higher and higher mass scales by increasing the collision energy and accumulating more data.
On the other hand, all the measurements for the discovered particle agree with the predictions of the SM Higgs boson within the experimental uncertainty so far~\cite{Khachatryan:2016vau}. 
Such situation highly motivates us a thorough study of the Higgs sector at the LHC Run-II
as well as in future high-precision experimental programs such as the high-luminosity LHC (HL-LHC)~\cite{ATLAS:2013hta,CMS:2013xfa} 
and future lepton colliders such as the international linear collider (ILC)~\cite{Baer:2013cma,Asai:2017pwp,Fujii:2017vwa}, 
the lepton collision option of the future circular collider (FCC-ee)~\cite{Gomez-Ceballos:2013zzn}, 
the circular electron positron collider (CEPC)~\cite{CEPC-SPPCStudyGroup:2015csa} and 
the compact linear collider (CLIC)~\cite{CLIC:2016zwp}.

Although the SM takes the minimal setup for the Higgs sector; i.e., a scalar isospin doublet field,
there is no compelling reason to be minimal, and indeed many NP models predict additional scalar multiplets.
For instance, the $B-L$ extended SM~\cite{Khalil:2006yi} contains an additional singlet scalar field, 
while the minimal supersymmetric SM has another doublet scalar field.
Furthermore, some of the scenarios for radiative seesaw models predict extended Higgs sectors~\cite{Zee:1980ai,Zee:1985rj,Zee:1985id,Babu:1988ki,Krauss:2002px,Ma:2006km,Aoki:2008av},
and many of the scenarios of electroweak baryogenesis also require non-minimal structures for the Higgs sector~\cite{Turok:1990in,Bochkarev:1990gb,Nelson:1991ab}. 
Therefore, by studying the structure of the Higgs sector by experiments, we may be able to determine the models of NP.

There are two important consequences in models with a non-minimal Higgs sector.
One is the existence of additional scalar states, 
and 
the other is deviations of the interactions for the SM-like Higgs boson from the SM prediction due to the mixing with other neutral scalars as well as loop effects of additional scalars.
The current LHC Higgs program clearly targets them~\cite{deFlorian:2016spz},
and has been seeking for extra scalars in the direct searches~\cite{CMS:2016qbe,Aaboud:2017cxo,Aaboud:2017rel,Aaboud:2017gsl} and
looking for deviations in the Higgs coupling measurements~\cite{Khachatryan:2016vau,Aad:2015pla,CMS:2016qbe}, 
which already put constraints on the parameter space of extended Higgs models.
In this work, we focus on the latter aspect; i.e., deviations from the SM via precise measurements at the LHC as well as in the future experiments, 
where more accurate theoretical predictions are required not only in the SM but also in NP models.
Especially, once a deviation from the SM is detected, accurate NP predictions including loop effects become crucial to identify a model from the other models.

As the simplest extensions of the SM Higgs sector with the $\rho$ parameter to be unity at the tree level, we consider 
the Higgs singlet model (HSM) and the two Higgs doublet model (THDM).
In particular, we study the model with a real singlet scalar as the HSM, and four types of the THDM with softly broken $Z_2$ symmetry to avoid dangerous flavor changing neutral currents.

In Ref.~\cite{Kanemura:2017gbi} a full set of the numerical code (\hcoup\!\!) to evaluate the renormalized gauge invariant vertex functions for the SM-like Higgs boson
has been released 
in various extended Higgs models: the HSM, four types of the THDM and the inert doublet model (IDM), 
based on Refs.~\cite{Kanemura:2004mg,Kanemura:2014dja,Kanemura:2015mxa,Kanemura:2015fra,Kanemura:2016sos,Kanemura:2016lkz,Kanemura:2017wtm}.
However, 
in order to compare the theory predictions on the Higgs boson couplings with higher-order calculations with experimental data,
we should directly evaluate physics quantities such as production cross sections and decay branching ratios
instead of, for example, the renormalized vertex functions or something like the $\kappa$ parameters (the scale factors) for the Higgs boson couplings which might not be well defined beyond the leading order (LO). 

In this letter, 
in order to identify the Higgs sector using future precision data, 
we calculate the partial decay widths of the discovered Higgs boson ($h$)
into fermion pairs and gauge-boson pairs
with one-loop electroweak (EW) and one-loop QCD corrections  
in the HSM and four types of the THDM.
So far, 
a few next-to-leading order (NLO) EW radiative corrections in the THDMs have been done for $h\to bb/\tau\tau$~\cite{Arhrib:2003ph,Arhrib:2016snv} and for $h\to ZZ^*\to Z\ell\ell$~\cite{Castilla-Valdez:2015sng}.
NLO EW and QCD calculations for $h\to WW/ZZ\to 4\,$fermions
in the 
HSM~\cite{Altenkamp:2018bcs} as well as in the THDMs~\cite{Altenkamp:2017ldc,Altenkamp:2017kxk} were also reported.
However, there has been no study to compute all the decay rates with higher-order corrections in various extended Higgs models comprehensively.
We employ and extend \hcoup to systematically compute all the decay rates of the SM-like Higgs boson in each model at the one-loop level.
In order to study the deviations from the SM predictions, 
we evaluate the ratios of the decay rates in extended Higgs models to those in the SM. 
We then
discuss how the NP model can be identified by the pattern of the deviations.
Furthermore, we study how we can obtain information on the mass of the additional Higgs bosons from detecting the deviations in the decay rates of the Higgs boson.

From the current LHC data, the Higgs couplings to the weak gauge bosons have been measured 
with the precision of about 10\% level,
and those to $\tau\tau$ and $\gamma\gamma$ ($bb$) are about 20 (50)\% at 1$\sigma$~\cite{Khachatryan:2016vau}.
The accuracy for the measurements of the Higgs couplings is expected to be improved in future experiments. 
For instance, 
the uncertainties will be down to 
9\% 11\%, 4\% and 4.2\% for the Higgs couplings to $\tau\tau$, $bb$, $ZZ$ and $\gamma\gamma$ at the HL-LHC with the integrated luminosity of 3 ab$^{-1}$~\cite{ATL-PHYS-PUB-2014-016}.  
Furthermore, those will be  
1.9\% 1.8\%, 2.4\%, 0.38\% and 11\% for the Higgs couplings to $\tau\tau$, $bb$, $cc$, $ZZ$ and $\gamma\gamma$ at the ILC with the integrated luminosity of 2 ab$^{-1}$ at $\sqrt{s}=250$~GeV~\cite{Fujii:2017vwa}.  
Therefore, the radiative corrections to the Higgs boson decay rates discussed in this work are inevitable to be compared with these precise measurements in the future experiments. 

This letter is organized as follows.
In Sec.~\ref{sec:models} we introduce extended Higgs models which we consider, namely the HSM and the THDM.
We describe the framework of the calculation for the Higgs decay rates in Sec.~\ref{sec:hwidth}, 
and show numerical results in Sec.~\ref{sec:results}.
The conclusion is given in Sec.~\ref{sec:summary}.
 
\section{Extended Higgs models}\label{sec:models}

We briefly describe the HSM and the THDM in order. 
We give the Higgs potential, and define the input parameters for each model.

\subsection{Higgs singlet model}
\label{HSM}

In addition to an isospin doublet Higgs field $\Phi$ with hypercharge $Y=1/2$ as in the SM,
the HSM has a real singlet scalar field $S$ with $Y=0$. 
The most general Higgs potential is written as~\cite{Chen:2014ask,Kanemura:2015fra}
\begin{align}
V(\Phi,S) =&\, m_\Phi^2|\Phi|^2+\lambda |\Phi|^4  
+\mu_{\Phi S}^{}|\Phi|^2 S+ \lambda_{\Phi S} |\Phi|^2 S^2 
+t_S^{}S +m^2_SS^2+ \mu_SS^3+ \lambda_SS^4,
\label{Eq:HSM_pot}
\end{align}
where all the parameters are real. 
The scalar fields $\Phi$ and $S$ are expressed in terms of the component fields by 
\begin{align}
\Phi=\left(\begin{array}{c}
G^+\\
\frac{1}{\sqrt{2}}(v+\phi+iG^0)
\end{array}\right),\quad
S=v_S^{} + s,
\label{hsm_f}
\end{align}
where $v$ and $v_S$ are the vacuum expectation value (VEV), and $G^{\pm,0}$ are the Nambu-Goldstone bosons to be absorbed by the longitudinal components of the weak gauge bosons.
The potential is invariant under the shift of the VEV for the singlet field, so that $v_S$ can be fixed to be zero without any loss of generality~\cite{Chen:2014ask}.

The mass eigenstates of the Higgs bosons are defined by introducing the mixing angle as
\begin{align}
\begin{pmatrix}
s \\
\phi
\end{pmatrix} = R(\alpha)
\begin{pmatrix}
H \\
h
\end{pmatrix}~~\text{with}~~R(\alpha) = 
\begin{pmatrix}
c_\alpha & -s_ \alpha \\
s_\alpha & c_\alpha
\end{pmatrix},   \label{mat_r}
\end{align}
where $s_\theta$ and $c_\theta$ represent $\sin\theta$ and $\cos\theta$, respectively. 
The squared masses and the mixing angle are expressed as
\begin{align}
 &m_H^2=M_{11}^2c^2_\alpha +M_{22}^2s^2_\alpha +M_{12}^2s_{2\alpha},\quad 
 m_h^2=M_{11}^2s^2_\alpha +M_{22}^2c^2_\alpha -M_{12}^2s_{2\alpha}, \notag \\
 &\tan 2\alpha=\frac{2M_{12}^2}{M_{11}^2-M_{22}^2}, \label{tan2a}
\end{align}  
where the mass matrix elements $M^2_{ij}$ are given by
\begin{align}
M^2_{11}= M^2+ v^2\lambda_{\Phi S}  ,\quad M^2_{22}=2v^2\lambda,\quad M^2_{12}=v\mu_{\Phi S}^{}, 
\label{mij}
\end{align}
with $M^2\equiv 2m_S^2$.
The parameters $m_\Phi^2$ and $t_S^{}$ are eliminated by using the stationary conditions for $\phi$ and $s$. 

There are seven parameters in the Higgs potential, which are rewritten by  
\begin{align}
 v,~m_h,~m_H,~M^2,~\mu_{S},~\lambda_S,~\alpha,
\end{align}
among which $v$ and $m_h$ are fixed to be $(\sqrt{2} G_F)^{-1/2}\simeq 246~{\rm GeV}$ 
with $G_F$ being the Fermi constant
and 125~GeV, respectively.
The rest five parameters are free parameters of the model.

The size of the dimensionless parameters in the potential can be constrained by imposing 
bounds from perturbative unitarity and vacuum stability. 
These constraints are translated in terms of the constraints on physical quantities, i.e., masses of Higgs bosons and mixing angles via the relations given in Eqs.~(\ref{tan2a})--(\ref{mij}).  
Concerning the former bound, all the independent eigenvalues of the $s$-wave amplitude matrix for the elastic $2\to 2$ body scatterings have been derived in Ref.~\cite{Cynolter:2004cq}. 
For the latter bound, the necessary conditions to guarantee the potential to be bounded from below at large field values has been given in Ref.~\cite{Pruna:2013bma}. 
In addition to these bounds, one has to avoid wrong local extrema in which the true vacuum, $\langle\Phi^0\rangle = v/\sqrt{2}$, does not correspond to the deepest minimum. 
Such wrong vacuum can appear due to the existence of the scalar trilinear couplings $\mu_{\Phi S}^{}$ and $\mu_{S}^{}$. 
In Refs.~\cite{Espinosa:2011ax,Chen:2014ask,Lewis:2017dme}, the conditions to avoid the wrong vacuum have been found. 

It is also important to take into account the constraint from EW oblique parameters such as the $S$ and $T$ parameters~\cite{Peskin:1990zt}. 
These parameters are calculated in terms of the gauge boson 2-point functions whose analytic formulae can be found in Ref.~\cite{Lopez-Val:2014jva}. 
In addition to the above constraints, LHC data, i.e., 
direct searches for additional Higgs bosons and
signal strengths for the discovered Higgs boson can also set 
a limit on the mass of the extra Higgs boson and their couplings to SM particles.  
In the HSM, constraints on $m_H^{}$ and $\alpha$ have been studied by using LHC Run-I~\cite{Robens:2015gla} and Run-II data~\cite{Robens:2016xkb, Blasi:2017zel,Gu:2017ckc}.

\subsection{Two Higgs doublet model}

Instead of a singlet field as in the HSM, the THDM has an additional isospin doublet scalar field with $Y=1/2$.
In order to avoid flavor changing neutral currents at the tree level,
we impose a $Z_2$ symmetry~\cite{Glashow:1976nt}, which can be broken softly. 
Under this symmetry the Higgs potential is given by
\begin{align}
V(\Phi_1,\Phi_2) &= m_1^2|\Phi_1|^2+m_2^2|\Phi_2|^2-m_3^2(\Phi_1^\dagger \Phi_2^{} +\text{h.c.})\notag\\
&\quad +\frac{1}{2}\lambda_1|\Phi_1|^4+\frac{1}{2}\lambda_2|\Phi_2|^4+\lambda_3|\Phi_1|^2|\Phi_2|^2+\lambda_4|\Phi_1^\dagger\Phi_2^{}|^2
+\frac{1}{2}\lambda_5[(\Phi_1^\dagger\Phi_2^{})^2+\text{h.c.}],
\label{pot_thdm2}
\end{align}
where all the parameters can be real by assuming the $CP$ conservation.
The two doublet fields $\Phi_1$ and $\Phi_2$ are parameterized as 
\begin{align}
\Phi_i=\left(\begin{array}{c}
w_i^+\\
\frac{1}{\sqrt{2}}(v_i+h_i+iz_i)
\end{array}\right)~~\text{with}~~i=1,2,
\end{align} 
where $v_1$ and $v_2$ are the VEVs of the Higgs doublet fields with $v=(v_1^2+v_2^2)^{1/2}$.

The mass eigenstates of the Higgs fields are defined as follows:
\begin{align}
\left(\begin{array}{c}
h_1\\
h_2
\end{array}\right)=R(\alpha)
\left(\begin{array}{c}
H\\
h
\end{array}\right),
\quad
\left(\begin{array}{c}
z_1\\
z_2
\end{array}\right)
=R(\beta)\left(\begin{array}{c}
G^0\\
A
\end{array}\right),
\quad
\left(\begin{array}{c}
w_1^\pm\\
w_2^\pm
\end{array}\right)&=R(\beta)
\left(\begin{array}{c}
G^\pm\\
H^\pm
\end{array}\right),
\label{mixing}
\end{align}
where $\tan\beta = v_2/v_1$. 
By solving the two stationary conditions for $h_1$ and $h_2$, we can eliminate the parameters $m_1^2$ and $m_2^2$. 
Then, the squared masses of the physical Higgs bosons and the mixing angle $\alpha$ are expressed by 
\begin{align}
&m_{H^\pm}^2 = M^2-\frac{1}{2}v^2(\lambda_4+\lambda_5),\quad \notag
 m_A^2=M^2-v^2\lambda_5,  \\ 
&m_H^2=M_{11}^2 c^2_{\beta-\alpha} + M_{22}^2 s^2_{\beta-\alpha} -M_{12}^2 s_{2(\beta-\alpha)}, \quad 
m_h^2=M_{11}^2 s^2_{\beta-\alpha}  + M_{22}^2 c^2_{\beta-\alpha}  +M_{12}^2 s_{2(\beta-\alpha)}, \notag \\
&\tan 2(\beta-\alpha)= -\frac{2M_{12}^2}{M_{11}^2-M_{22}^2}, \label{333}
\end{align} 
where $M^2 \equiv m_3^2/s_\beta c_\beta$ describes the soft breaking scale of the $Z_2$ symmetry, and $M_{ij}^2$ are the mass matrix elements for the $CP$-even scalar states in the basis of $(h_1,h_2)R(\beta)$:
\begin{align}
&M_{11}^2=v^2(\lambda_1c^4_\beta+\lambda_2s^4_\beta+\lambda_{345}s^2_{\beta}c^2_\beta),\quad
M_{22}^2=M^2+\frac{1}{2}v^2s^2_{2\beta}(\lambda_1+\lambda_2-2\lambda_{345}), \notag\\
&M_{12}^2=\frac{1}{2}v^2 s_{2\beta}( -\lambda_1c^2_\beta+\lambda_2s^2_\beta+\lambda_{345} c_{2\beta}  ),
\end{align}
with $\lambda_{345}\equiv \lambda_3+\lambda_4+\lambda_5$. 
We choose the seven free parameters in the THDM:
\begin{align}
m_H^{},~ m_A^{},~ m_{H^\pm},~ M^2,~ \tan\beta,~ s_{\beta-\alpha}(\geq0),~ {\rm Sign}(c_{\beta-\alpha}).
\end{align} 

For the Yukawa sector, we can define four types of the interactions under the softly-broken $Z_2$ symmetry~\cite{Barger:1989fj,Grossman:1994jb,Aoki:2009ha},   
as shown in Table~\ref{yukawa_tab}, depending on the $Z_2$ charge assignment for the right-handed fermions. 
As seen later, the difference in the Yukawa sector plays an important role for the Higgs decay rates in each type of the THDM.  

\begin{table}
\begin{center}
{
\begin{tabular}{l|ccccccc|ccc}\hline
&$\Phi_1$&$\Phi_2$&$Q_L$&$L_L$&$u_R$&$d_R$&$e_R$&$\zeta_u$ &$\zeta_d$&$\zeta_e$ \\\hline
Type-I &$+$&
$-$&$+$&$+$&
$-$&$-$&$-$&$\cot\beta$&$\cot\beta$&$\cot\beta$ \\ 
Type-II&$+$&
$-$&$+$&$+$&
$-$
&$+$&$+$& $\cot\beta$&$-\tan\beta$&$-\tan\beta$ \\ 
Type-X (lepton-specific) &$+$&
$-$&$+$&$+$&
$-$
&$-$&$+$&$\cot\beta$&$\cot\beta$&$-\tan\beta$ \\ 
Type-Y (flipped) &$+$&
$-$&$+$&$+$&
$-$
&$+$&$-$& $\cot\beta$&$-\tan\beta$&$\cot\beta$ \\\hline
\end{tabular}}
\caption{The $Z_2$ charge assignment and the $\zeta_f$ factors in Eq.~\eqref{kappa_thdm} for each type of the Yukawa interactions. 
}
\label{yukawa_tab}
\end{center}
\end{table}

As we mentioned in the previous subsection, 
regions of the parameter space can be constrained by imposing the bounds from 
the perturbative unitarity~\cite{Kanemura:1993hm,Akeroyd:2000wc,Ginzburg:2005dt,Kanemura:2015ska}, the vacuum stability~\cite{Deshpande:1977rw,Sher:1988mj,Kanemura:1999xf}
and the $S$, $T$ parameters~\cite{Toussaint:1978zm,Bertolini:1985ia,Peskin:2001rw,Grimus:2008nb,Kanemura:2011sj}. 
Constraints from the LHC Run-I and Run-II data have been discussed in Refs.~\cite{Bernon:2015qea,Dorsch:2016tab,Han:2017pfo,Arbey:2017gmh,Chang:2015goa,Blasi:2017zel,Gu:2017ckc}.
Differently from the HSM, in the THDMs flavor experiments such as $B$ meson decays also give an important constraint particularly on the mass of the charged Higgs bosons and $\tan\beta$. 
Constraints from various $B$ physics processes have been studied comprehensively in Refs.~\cite{Mahmoudi:2009zx,Enomoto:2015wbn}.  

\section{Decay widths of the SM-like Higgs boson at one loop}\label{sec:hwidth}

In this section we describe the framework to calculate the Higgs decay rates at one loop.

\subsection{Higgs boson vertices} 

We begin with definitions of the renormalized Higgs boson vertices $hff$ and $hVV$, 
which are the main piece of the one-loop calculation for the Higgs decay rates and 
can be evaluated by \hcoup\cite{Kanemura:2017gbi}.
We apply the improved on-shell renormalization scheme adopted in Refs.~\cite{Fleischer:1980ub,Krause:2016oke,Kanemura:2017wtm}, where
gauge dependence appearing in the renormalization of mixing angles among scalar fields is removed by using the pinch technique.
We calculate the one-loop amplitudes in the Feynman gauge. 
The renormalized $hff$ and $hVV$ vertices can be decomposed by the following form factors: 
\begin{align}
\hat{\Gamma}_{h ff}(p_1^2,p_2^2,q^2)&=
\hat{\Gamma}_{h ff}^S+\gamma_5 \hat{\Gamma}_{h ff}^P+p_1\hspace{-3.5mm}/\hspace{2mm}\hat{\Gamma}_{h ff}^{V_1}
+p_2\hspace{-3.5mm}/\hspace{2mm}\hat{\Gamma}_{h ff}^{V_2}\notag\\
&\quad +p_1\hspace{-3.5mm}/\hspace{2mm}\gamma_5 \hat{\Gamma}_{h ff}^{A_1}
+p_2\hspace{-3.5mm}/\hspace{2mm}\gamma_5\hat{\Gamma}_{h ff}^{A_2}
+p_1\hspace{-3.5mm}/\hspace{2mm}p_2\hspace{-3.5mm}/\hspace{2mm}\hat{\Gamma}_{h ff}^{T}
+p_1\hspace{-3.5mm}/\hspace{2mm}p_2\hspace{-3.5mm}/\hspace{2mm}\gamma_5\hat{\Gamma}_{h ff}^{PT},\\ 
\hat{\Gamma}_{h VV}^{\mu\nu}(p_1^2,p_2^2,q^2)&=g^{\mu\nu}\hat{\Gamma}_{h VV}^1
+\frac{p_1^\nu p_2^\mu}{m_V^2}\hat{\Gamma}_{h VV}^2
+i\epsilon^{\mu\nu\rho\sigma}\frac{p_{1\rho} p_{2\sigma}}{m_V^2}\hat{\Gamma}_{h VV}^3,  \label{form_factor}
\end{align}
where $p_i^{\mu}$ is the incoming momentum of a fermion or a vector boson and $q^\mu$ is the momentum of the Higgs boson. 
These renormalized form factors $\hat{\Gamma}^i_{hXX}$ are expressed by
\begin{align}
\hat{\Gamma}^i_{hXX}(p_1^2,p_2^2,q^2)&=\Gamma^{i,{\rm tree}}_{hXX}+\Gamma^{i,{\rm loop}}_{hXX}
 =\Gamma^{i,{\rm tree}}_{hXX}+\Gamma^{i,{\rm 1PI}}_{hXX}(p_1^2,p_2^2,q^2)+\delta \Gamma^{i}_{hXX},
\label{eq:renohVV} 
\end{align}
where $\Gamma_{hXX}^{i,{\rm tree}}$, $\Gamma_{hXX}^{i,{\rm 1PI}}$ and $\delta\Gamma^i_{hXX}$ denote 
the contributions from the tree-level diagram, 1PI diagrams for the vertex and the counterterms, respectively. 

The tree-level contributions are expressed as
\begin{align}
 \Gamma_{hff}^{S,{\rm tree}}=-\frac{m_f}{v}\kappa_f,\quad
 \Gamma_{hVV}^{1,{\rm tree}}=\frac{2m_V^2}{v}\kappa_V,
\end{align}
where the scaling factors $\kappa_f$ and $\kappa_V$ in each model are given by
\begin{align}
 &\kappa_f=\kappa_V=c_\alpha \quad \text{in\ the\ HSM},
 \label{kappa_hsm}\\
 &\kappa_f=s_{\beta-\alpha}+\zeta_f c_{\beta-\alpha}, \quad \kappa_V=s_{\beta-\alpha}
 \quad\text{in\ the\ THDMs}.
 \label{kappa_thdm}
\end{align}
The mixing parameters in each model are defined in Sec.~\ref{sec:models}, where $\zeta_f$ is shown in Table~\ref{yukawa_tab}.
We note that the tree-level contributions to all the other form factors are zero.
Explicit formulae for $\Gamma_{hXX}^{i,{\rm 1PI}}$ in the HSM and the THDMs are presented in Refs.~\cite{Kanemura:2015fra} and \cite{Kanemura:2015mxa} 
respectively, 
and those for $\delta\Gamma^i_{hXX}$ in each model are given in Ref.~\cite{Kanemura:2017wtm}. 

\subsection{$h\to ff$}

We parametrize NLO corrections to the $h\to ff$ decay rate as
\begin{align}
 \Gamma(h\to ff)=\Gamma_0(h\to ff)[1+\Delta_{\rm weak}^f+\Delta_{\rm QED}^f+\Delta_{\rm QCD}^f],
\end{align}
where the weak and QED corrections are decomposed in order to treat the infrared divergence in the QED correction separately.  

The decay rate at the LO is given by
\begin{align}
 \Gamma_0(h\to ff)
 =\frac{\beta_f}{16\pi m_h}|{\cal M}^{\rm tree}_{hff}|^2
 =\frac{N_c m_h \beta_f^3}{8\pi}|\Gamma^{S,{\rm tree}}_{hff}|^2
 =\frac{N_c G_F m_f^2 m_h \beta_f^3}{4\sqrt{2}\pi}\kappa_f^2
 \label{G0_hff}
\end{align}
with $N_c=3\ (1)$ for quarks (leptons) and $\beta_f=(1-4m_f^2/m_h^2)^{1/2}$. 

The weak correction is expressed as~\cite{Kniehl:1991ze,Dabelstein:1991ky}
\begin{align}
 \Delta_{\rm weak}^f= -\Delta r
 +\frac{1}{|{\cal M}^{\rm tree}_{hff}|^2}
 2{\rm Re}({\cal M}^{\rm tree}_{hff}{\cal M}^{\rm loop*}_{hff}),
\label{weak_hff}
\end{align}
where $\Delta r$ is the weak corrections to the muon decay (see Appendix B in Ref.~\cite{Kanemura:2015fra} for the explicit formula).
The one-loop contribution ${\cal M}^{\rm loop}_{hff}$ to the amplitudes is written in terms of the one-loop part of the renormalized form factors defined above as
\begin{align}
 {\cal M}^{\rm loop}_{hff}&=m_h\beta_f\Big[
 \Gamma_{hff}^{S,{\rm loop}}
 +m_f(\Gamma^{V_1,{\rm loop}}_{hff}-\Gamma^{V_2,{\rm loop}}_{hff})
 +(m_h^2-m_f^2)\Gamma_{hff}^{T,{\rm loop}} \Big],
\end{align}
where we fix the momenta of the form factors as $p_1^2 = p_2^2 = m_f^2$ and $q^2 = m_h^2$.
We note that with these momenta
$\Gamma^{V_1,{\rm loop}}_{hff}=-\Gamma^{V_2,{\rm loop}}_{hff}$ and 
the form factors proportional to $\gamma_5$ do not contribute to the decay rate. 

The QED correction comes from the photon-exchange loop and the corresponding counterterm as well as the real photon emission, and is common to the SM~\cite{Bardin:1990zj,Kniehl:1991ze,Dabelstein:1991ky},
which is given in the on-shell renormalization scheme by
\begin{align}
 \Delta_{\rm QED}^f=
 \frac{\alpha_{\rm EM}}{\pi}Q_f^2\Big[\frac{9}{4}+\frac{3}{2}\log\Big(\frac{m_f^2}{m_h^2}\Big)\Big]
\end{align}
in the limit of $m_f^2\ll m_h^2$,%
\footnote{We keep $\beta_f$ in the numerical computation; see Ref.~\cite{Dabelstein:1991ky} for the explicit formula.} 
where $Q_f$ is the electric charge of the fermions.

For the Higgs boson decay into a quark pair, the QCD correction is dominant over the EW correction and known up to order $\alpha_s^4$ in the SM; see, e.g. a recent paper~\cite{Mihaila:2015lwa}, where the mixed QCD-EW (${\cal O}(\alpha_s\alpha_{\rm EM})$) corrections are also presented.
Similar to the QED correction, the QCD correction is common between the SM and the extended Higgs models.
In our calculation, we include the one-loop QCD correction%
\footnote{As long as we focus on the deviations from the SM predictions, one-loop corrections are sufficient.}
in the $\overline{\rm MS}$ renormalization scheme
\begin{align}
 \Delta_{\rm QCD}^f=
\frac{\bar\alpha_s(\mu)}{\pi}C_F\Big[\frac{17}{4}+\frac{3}{2}\log\Big(\frac{\mu^2}{m_h^2}\Big)\Big],
\end{align}
where $C_F=4/3$ and we choose $\mu^2=m_h^2$.
The pole mass for the quarks in Eq.~\eqref{G0_hff} should be replaced by the running mass $\bar m_f(\mu)$, and $\bar m_f$ and $\bar\alpha_s$ are defined at the scale $m_h$.
We also adopt the $\overline{\rm MS}$ scheme for the QED correction to the decay rates into a quark pair, which is obtained from $\Delta_{\rm QCD}^f$ by the replacement $\bar\alpha_s\to Q_f^2\alpha_{\rm EM}$ and $C_F\to1$.

\subsection{$h\to VV^*$}
\label{sec:hvv}

The 125~GeV Higgs boson can also decay into a pair of weak bosons with one on-shell $V$ and the other off-shell $V^*$. 
In this work we evaluate the NLO EW and QCD corrections to the three-body decay $h\to Vff$ to treat the off-shell gauge boson properly.

For $V=Z$, similar to $h\to ff$, 
we parametrize the one-loop corrections to the $h\to Zff$ decay rate as
\begin{align}
 \Gamma(h\to Zff)=\Gamma_0(h\to Zff)[1+\Delta_{\rm weak}^Z+\Delta_{\rm QED}^Z+\Delta_{\rm QCD}^Z].
\end{align}
At the LO the $h\to Zff$ decay only happens via the off-shell $Z$ boson, 
$h\to ZZ^*\to Zff$,
and the decay rate is given by~\cite{Pocsik:1980ta,Rizzo:1980gz,Keung:1984hn}
\begin{align}
 \Gamma_0(h\to Zff)=
 \frac{1}{2m_h}\int d\Phi_3\,|{\cal M}^{\rm tree}_{hZff}|^2 
 =\frac{G_F^2m_Z^4m_h}{24\pi^3}(v_f^2+a_f^2)F\Big(\frac{m_Z^2}{m_h^2}\Big)\kappa_V^2,
\end{align}
where 
$v_f=I_f/2-Q_f\sin\theta_{W}^2$ and $a_f=I_f/2$ are the vector and axial-vector couplings for the weak neutral-current interactions 
with the isospin of the fermions $I_f$ and the weak mixing angle $\theta_W$.
The fermion mass is ignored in this expression.
The function $F(x)$ is given by
\begin{align}
 F(x)=\frac{3(1-8x+20x^2)}{(4x-1)^{1/2}}\arccos\Big(\frac{3x-1}{2x^{3/2}}\Big)
 -\frac{1-x}{2x}(2-13x+47x^2)-\frac{3}{2}(1-6x+4x^2)\log x.
\end{align}

\begin{figure}
 \center 
 \includegraphics[width=0.9\textwidth]{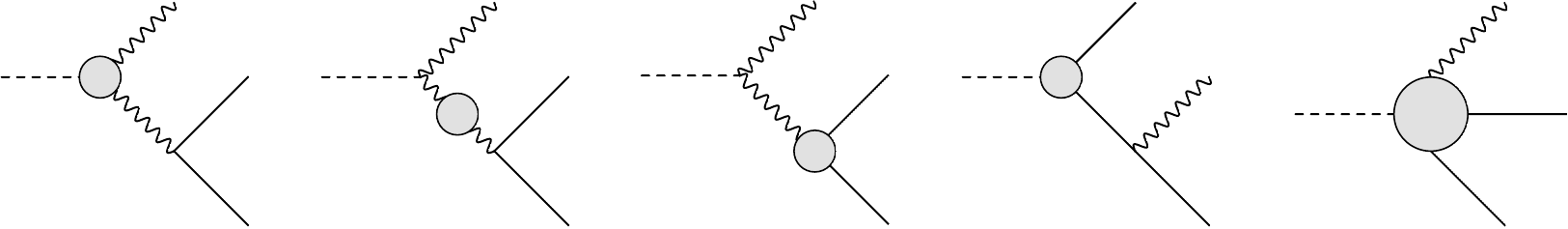}\\[-2mm] 
 (a)\hspace*{26.25mm}(b)\hspace*{26.25mm}(c)\hspace*{26.25mm}(d)\hspace*{26.25mm}(e)
 \caption{Schematic diagrams of each loop contribution to $h\to Vff$.}
\label{fig:diagram}
\end{figure}

The weak correction is given by~\cite{Kniehl:1991hk,Denner:1992bc}
\begin{align}
 \Delta_{\rm weak}^Z= -2\Delta r-\Delta Z_{\rm wf}
 +\int d\Phi_3\frac{1}{|{\cal M}^{\rm tree}_{hZff}|^2}
 2{\rm Re}({\cal M}^{\rm tree}_{hZff}{\cal M}^{\rm loop*}_{hZff}),
\end{align}
where $\Delta r$ is the same as in Eq.~\eqref{weak_hff} and
$\Delta Z_{\rm wf}$ is due to the $Z$-boson wavefunction renormalization.
The one-loop contribution ${\cal M}^{\rm loop}_{hZff}$ to the amplitudes is the sum of each loop contribution whose diagram is schematically depicted in order in Fig.~\ref{fig:diagram}\,(a)-(e):  
\begin{align} 
  {\cal M}^{\rm loop}_{hZff}={\cal M}^{\rm loop}_{a}+{\cal M}^{\rm loop}_{b}
 +{\cal M}^{\rm loop}_{c}+{\cal M}^{\rm loop}_{d}+{\cal M}^{\rm loop}_{e}.
\end{align}
The amplitude ${\cal M}^{\rm loop}_{a}$ represented by Fig.~\ref{fig:diagram}\,(a)
contains the one-loop part of the renormalized vertices $\Gamma^{i,{\rm loop}}_{hVV}$, 
while ${\cal M}^{\rm loop}_{b}$ by Fig.~\ref{fig:diagram}\,(b) 
has the renormalized one-loop gauge-boson 2-point functions.  
Additional scalars can contribute in these loops. 
For $m_f=0$,
the extended Higgs sectors do not affect the $Zff$ vertex in the diagram (c) in Fig.~\ref{fig:diagram},
but modify the $hff$ vertex in the diagram (d) and the box diagram (e) only by the mixing effects 
as the $\kappa_V$ factor. 
The renormalized vertices $\Gamma^{i,{\rm loop}}_{hVV}$ and $\Gamma^{i,{\rm loop}}_{hff}$ 
as well as the renormalized 2-point functions can be evaluated by using \hcoup\!\!,
while
we newly implemented the renormalized $Zff$ vertices in SM and the $hZff$ box contributions in the {\sc H-COUP} framework.
Numerical computations were done in two ways independently, with the trace technique and with the helicity amplitude calculation, and confirmed to reproduce the previous works in the SM~\cite{Kniehl:1991hk,Denner:1992bc} and to agree with each other in extended Higgs models, 
which will be reported elsewhere with all the explicit formulae~\cite{Kanemura:2018xxx}.

The QED and QCD corrections come from photon and gluon exchanges between the fermions in the final state as well as real emissions.  
They are taken into account by~\cite{Kniehl:1993ay}
\begin{align}
 \Delta_{\rm QED}^Z+\Delta_{\rm QCD}^Z=\frac{3}{4\pi}(Q_f^2\alpha_{\rm EM} + C_F\bar\alpha_s).
\end{align}

The decay rate $\Gamma(h\to Wff')$ can be calculated in analogy to the case of $V=Z$ discussed above.
Differences are that $W$ bosons can emit a photon and the separation between weak and QED corrections cannot be done. 
The study of this decay mode will be presented in Ref.~\cite{Kanemura:2018xxx}.

We note that, in this way of the calculation for the Higgs decays into weak-boson pairs,   
we cannot take into account the interference of some specific final state in the Higgs decays into four fermions such as $h\to e^+e^-e^+e^-$, $e^+\nu_ee^-\bar\nu_e$, etc,
whose contributions to the total width in the SM is very small, about 0.2\%~\cite{Bredenstein:2006rh,deFlorian:2016spz}.

\subsection{$h\to gg/\gamma\gamma/Z\gamma$}

\hcoup can provide the decay rates of the loop-induced processes $h\to gg/\gamma\gamma/Z\gamma$ in extended Higgs models at LO.
Their explicit analytic formulae in the HSM and the THDMs are given in Refs.~\cite{Kanemura:2015fra} and \cite{Kanemura:2015mxa}, respectively. 
The QCD corrections are taken into account in the heavy top-quark limit; see, e.g., Ref.~\cite{Djouadi:2005gi}.
 
\section{Numerical evaluations}\label{sec:results}

\subsection{Parameter spaces}

In order to discuss deviations from predictions in the SM, we evaluate the ratios of the partial decay rates
\begin{align}
 \Delta R(h\to XX)=\frac{\Gamma(h\to XX)}{\Gamma_{\rm SM}(h\to XX)}-1,
\label{delR}
\end{align} 
where $\Gamma_{\rm SM}(h\to XX)$ is the partial decay rate in the SM with the one-loop EW and one-loop QCD corrections.
Assuming the discovered Higgs boson with the mass of 125~GeV as the lightest Higgs boson $h$,
we take several sets for masses of the additional Higgs bosons, and we scan the other internal model parameters in each model in the following way.

In the HSM, there are five free parameters; i.e., $m_H$, $c_\alpha$, $M^2$, $\mu_S$ and $\lambda_S$. 
The mass of the second Higgs boson $H$ is taken as 
\begin{align}
 &m_H=500, 1000, 2000, 3000\ {\rm and}\ 5000\GeV,
 \label{mass_hsm}
\end{align}
while $c_\alpha$ and $M^2$ are scanned as 
\begin{align}
 &0.95<c_\alpha<1,\quad 
 0<M^2<m_H^2.
 \label{sca_HSM}
\end{align}
Here, the region of $c_\alpha$ is taken so that the model is not far from the SM-like limit. 
The other parameters $\mu_S$ and $\lambda_S$ are taken to be zero, 
because
the $\mu_S$ dependence in $\Delta R$'s is numerically negligible
while $\lambda_S$ is irrelevant to our current analysis.

In the THDMs, there are seven free parameters; i.e., $m_H$, $m_A$, $m_{H^\pm}$, $s_{\beta-\alpha}$, $\tan\beta$, $M^2$ and Sign($c_{\beta-\alpha}$).
In order to avoid the constraint from the $T$ parameter, 
we take $m_{H^{\pm}}=m_A$ by which new contributions to the $T$ parameter is suppressed due to the custodial symmetry~\cite{Haber:1992py,Pomarol:1993mu}.  
Furthermore, throughout our analysis we assume $m_H=m_{H^{\pm}}=m_A$ for simplicity. 
The degenerate mass is taken as 
\begin{align}
 &m_H=400,700,1000, 1500\ {\rm and}\ 2000\GeV.
 \label{mass_thdm}
\end{align} 
The other parameters are scanned as%
\footnote{The lower bound of $\tan\beta$ comes from the constraint of flavor experiments~\cite{Enomoto:2015wbn}.
For the upper limit of $\tan\beta$, a larger value can also be taken near the limit of $s_{\beta-\alpha}=1$,
where deviations in the decay rates are too small to be detected.
We do not consider the case where the Yukawa coupling constant changes the sign by a large value of $\tan\beta$ for $c_{\beta-\alpha}<1$,
because such a case has already been mostly excluded by the current LHC data~\cite{Aad:2015pla,CMS:2016qbe}.
}
\begin{align}
 &0.95<s_{\beta-\alpha}<1,\quad 1<\tan\beta<3,\quad
 0<M^2<m_H^2.
 \label{sca_THDM}
\end{align} 
Similar to $c_\alpha$ in the HSM, the region of $s_{\beta-\alpha}$ is taken to be close to the SM-like limit.
Finally, we consider both the positive and negative cases of $c_{\beta-\alpha}$.

Over the above parameter spaces, we take into account the following constraints discussed in Sec.~\ref{sec:models}: 
perturbative unitarity, vacuum stability, and the compatibility to the EW $S$ and $T$ parameters.
In addition, a condition to avoid wrong vacua is imposed to the parameter space in the HSM.
Those constraints are already implemented in \hcoup\!\!; see the manual~\cite{Kanemura:2017gbi} in details. 

We briefly mention the constraints from the LHC Higgs measurements. 
By using the data of approximately 5~fb$^{-1}$ at $\sqrt{s}=7$~TeV and 20~fb$^{-1}$ at $\sqrt{s}=8$~TeV, the ATLAS and CMS collaborations put the constraints on the mixing angles of the HSM and the THDMs as follows~\cite{Aad:2015pla,CMS:2016qbe}.
In the HSM, the observed (expected) limit at 95\% confidence level (CL) is $c_\alpha>0.94\ (0.88)$.
In the THDMs, the constraint on $\sba$ depends on the type of the models as well as $\tan\beta$.
For example, for $\tan\beta=1$ and $\cba>0$, the observed 95\% CL limits are $\sba>0.94$ in the Type-I THDM, $\sba>0.97$ in the Type-II and Type-Y THDMs, and $\sba>0.99$ in the Type-X THDM.
We note that those constraints are obtained by using the LO-motivated $\kappa$ framework~\cite{Heinemeyer:2013tqa}, 
and the interpretation of such constraints at the higher-order level might not be straightforward.

\subsection{$h\to\tau\tau$ vs. $h\to bb$}

\begin{figure}
 \center 
 \includegraphics[width=0.49\textwidth]{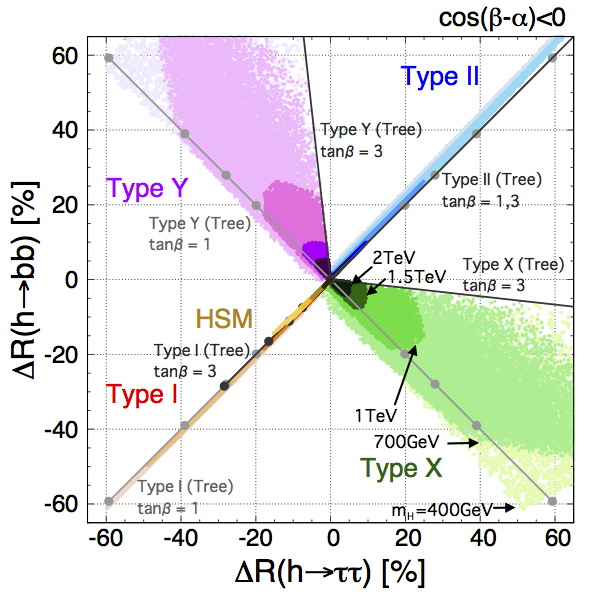}\ \
 \includegraphics[width=0.49\textwidth]{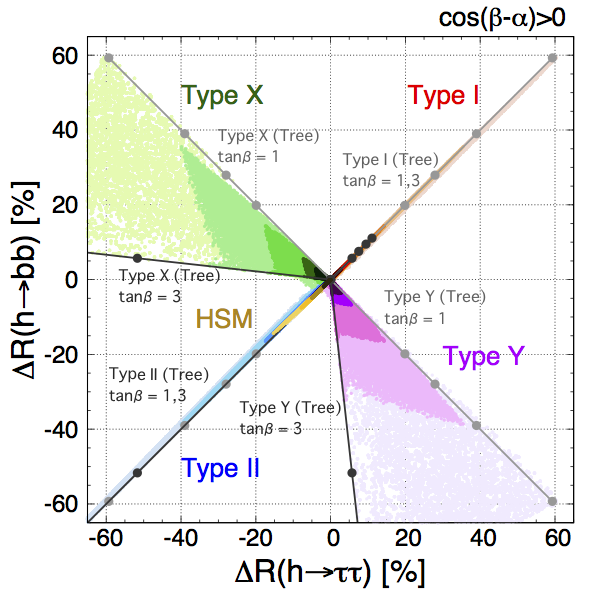}
 \caption{Correlation of the ratios of the Higgs decay widths in the HSM and four types of the THDM to those in the SM, defined in Eq.~\eqref{delR}, between $h\to\tau\tau$ and $h\to bb$. 
 The masses of additional Higgs bosons are taken as 500, 1000, 2000, 3000 and 5000~GeV in the HSM and as
 400, 700, 1000, 1500 and 2000~GeV in the THDMs.
 The left and right panels show the cases for $c_{\beta-\alpha}<0$ and $>0$ in the THDMs.
 The other parameters are scanned in the region shown in Eqs.~\eqref{sca_HSM} and \eqref{sca_THDM} 
 under the constraints described in the text,
 while $\mu_S$ and $\lambda_S$ in the HSM are taken to be zero. 
 As a reference, the tree-level predictions with $\tan\beta=1$ and 3 in the THDMs are also presented by gray and black lines, respectively. 
 }
\label{fig:dR_tb}
\end{figure}

In Fig.~\ref{fig:dR_tb} we show correlations of the ratios of the Higgs decay widths between $h\to\tau\tau$ and $h\to bb$
with one-loop EW and one-loop QCD corrections, defined in Eq.~\eqref{delR},
in the HSM and in four types of the THDM, i.e. Type-I, Type-II, Type-X and Type-Y.
The colored regions correspond to the predictions in these models; i.e.,
yellow, red, blue, green and purple correspond to the HSM, Type-I, Type-II, Type-X and Type-Y THDMs, respectively.
The contrast of the colors represents five mass scales of the additional Higgs bosons given in Eqs.~\eqref{mass_hsm} and \eqref{mass_thdm}
from light to dark. 
The evaluation is performed separately for $c_{\beta-\alpha}<0$ (left panel) and $>0$ (right panel) in the THDMs, while the other parameters are scanned in the regions shown in Eqs.~\eqref{sca_HSM} and \eqref{sca_THDM} under the constraints described above.
We also show the tree-level predictions with $\tan\beta=1$ and 3 in the THDMs by gray and black lines with dots denoting $s_{\beta-\alpha}=$1, 0.995, 0.99, 0.98, 0.95 from the origin,
 which corresponds to the SM limit.

The patterns of the deviations can be mainly determined by the tree-level mixing effects on the couplings, i.e. Eqs.~\eqref{kappa_hsm} and \eqref{kappa_thdm} in the HSM and the THDMs, respectively, 
which were studied 
in Ref.~\cite{Kanemura:2014bqa} in details.
In the HSM, as the mixing angle $c_\alpha$ is decreasing from the SM limit ($c_\alpha=1$), 
the decay widths are monotonically decreasing from the SM predictions 
both for the $\tau\tau$ and $bb$ modes.
On the other hand, those in each type of the THDM are quite distinctive due to the peculiar Yukawa structures. 
The dependence on the additional Higgs boson masses comes from the 
theoretical constraints such as perturbative unitarity and vacuum stability;
i.e., for a given mass of the heavy Higgs bosons the mixing parameters are constrained.
When the additional Higgs bosons are heavier by the growing $M^2$,
the mixing goes to zero and the model becomes closer to the SM-like limit.  
We note that the darker-colored regions include the lighter-colored regions.
In other words, when a deviation is observed, we can set the upper bound of the additional Higgs boson masses.

The above tree-level picture can be modified by quantum effects.
As expected, the behaviors of $\Delta R_{XX}(\equiv \Delta R(h\to XX))$ with radiative corrections are almost consistent with the analysis based on the $\kappa$ scheme in Refs.~\cite{Kanemura:2004mg,Kanemura:2014dja,Kanemura:2015mxa,Kanemura:2015fra},
but in the present analysis considerable improvements have been done in details of computations as already discussed. 
The pure quantum effects from the extra Higgs bosons can be observed, for example, in the Type-X THDM and the Type-Y THDM in
the region just below $\Delta R_{bb}=-\Delta R_{\tau\tau}$ for $c_{\beta-\alpha}<0$,
where those tend to reduce the decay widths. 
We note that in the Type-X THDM loop contributions to $\Gamma(h\to\tau\tau)$ are more sensitive to $\tb$ than those to $\Gamma(h\to bb)$~\cite{Kanemura:2014dja}.
On the contrary, 
in the Type-Y THDM loop contributions to $\Gamma(h\to bb)$ are more sensitive than those to $\Gamma(h\to\tau\tau)$.
Looking at Fig.~\ref{fig:dR_tb}(left), we see subtle differences in magnitudes of radiative corrections
between the Type-X THDM and the Type-Y THDM; namely, in the regions of the colored plots below the line of $\tb=1$ for the tree-level predictions.
The radiative corrections to the $h\to\tau\tau$ decay in the Type-X THDM is slightly larger than those to $h\to bb$ in the Type-Y THDM
even though the Yukawa structures for $h\tau\tau$ and $hbb$ are same.
This is due to the top-quark contributions to the $bb$ mode~\cite{Kanemura:2014dja}. 
From the detailed analyses with the fixed mixing angles,
which will be presented in details in elsewhere~\cite{Kanemura:2018xxx},
we find that 
the radiative corrections from the extended Higgs sector can be as large as several times per cent. 
Such large corrections are the consequence of the non-decoupling effect~\cite{Kanemura:2004mg,Kanemura:2014dja,Kanemura:2015mxa,Kanemura:2015fra} of the loop corrections
which is proportional to $m_H^2/(16\pi^2v^2)$ 
when $M^2\lesssim v^2$,
where the masses of the additional scalars mainly come from the VEV of the EW symmetry breaking; see the mass formulae in Sec.~\ref{sec:models}.
For a given mass of the additional Higgs boson(s), the minimum of $M^2$ determined by the theoretical constraints gives rise to the largest deviations.

\begin{figure}
 \center 
 \includegraphics[width=0.49\textwidth]{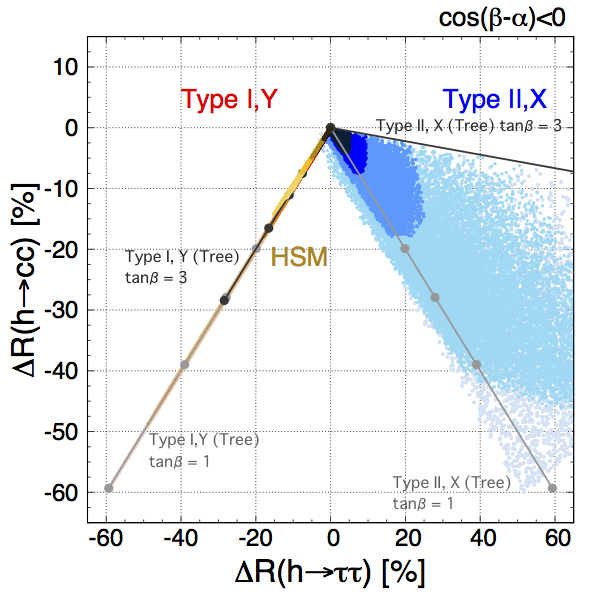}\ \
 \includegraphics[width=0.49\textwidth]{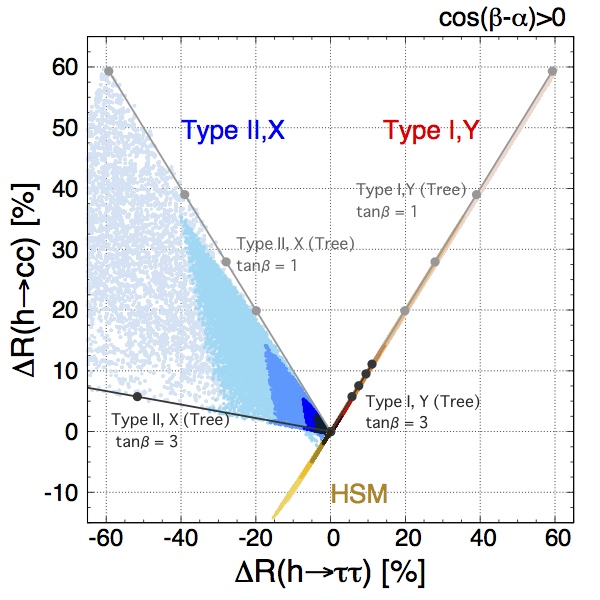}
 \caption{Correlation of the ratios of the Higgs decay widths in the HSM and four types of the THDM to those in the SM, defined in Eq.~\eqref{delR}, between $h\to\tau\tau$ and $h\to cc$. 
 See Fig.~\ref{fig:dR_tb} for the descriptions.}
\label{fig:dR_tc}
\end{figure}

\subsection{$h\to\tau\tau$ vs. $h\to cc$}

In Fig.~\ref{fig:dR_tc} we show the correlation in the radiative corrected decay rates of $h\to\tau\tau$ and $h\to cc$, 
where the descriptions for the figure are same as in Fig.~\ref{fig:dR_tb}.
At the tree level, 
the results in the Type-I (Type-II) THDM coincide with those in the Type-Y (Type-X) THDM, because the Yukawa structures for up-type quarks and leptons are common.
In contrast to the case in Fig.~\ref{fig:dR_tb},
the predicted region in the Type-II THDM spreads out,  
as the Yukawa structures between up-type quarks and leptons are different.
From the correlations among the three different fermionic decay modes of the Higgs boson shown in Figs.~\ref{fig:dR_tb} and \ref{fig:dR_tc}, 
we can identify a type of the THDM independently of the model parameters when a deviation is observed in experiments.

Similar to Fig.~\ref{fig:dR_tb}, quantum corrections of the additional Higgs bosons can be significant. 
For example, in the Type-II THDM for $c_{\beta-\alpha}<0$ we see 
the regions just below the line of the tree-level prediction at $\tb=1$.
The plots in these regions purely correspond to the quantum corrections.
We also find the dependence on the mass of the additional Higgs bosons, 
which is similar to that in the Type-X THDM in Fig.~\ref{fig:dR_tb}.

\subsection{$h\to\tau\tau$ vs. $h\to ZZ^*$}

\begin{figure}
 \center 
 \includegraphics[width=0.49\textwidth]{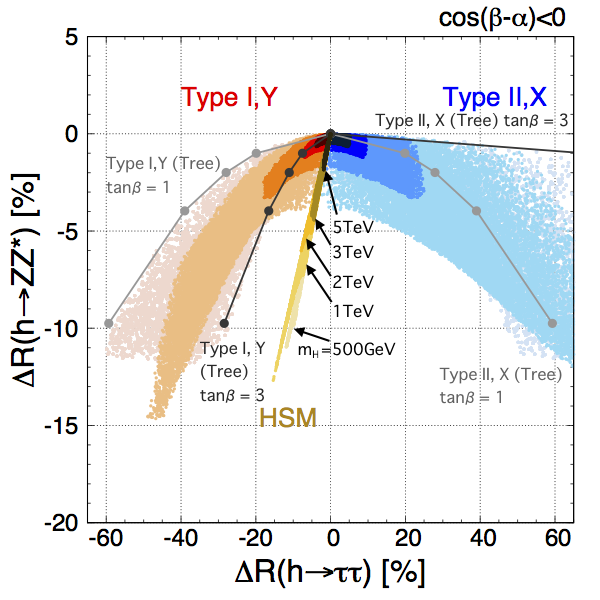}\ \
 \includegraphics[width=0.49\textwidth]{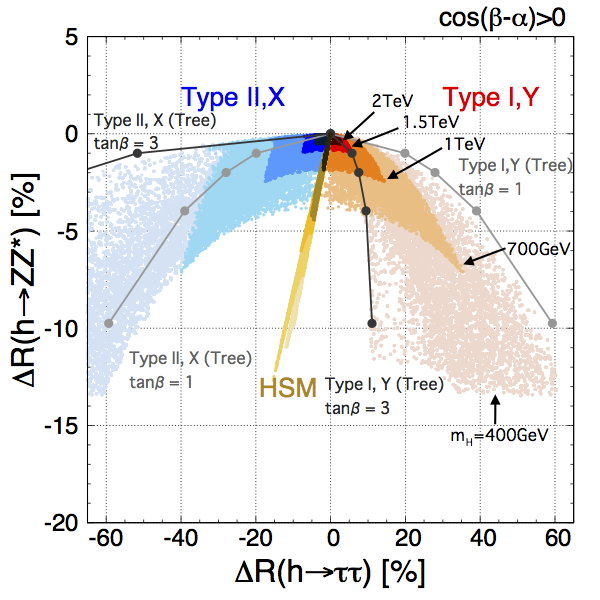}
 \caption{Correlation of the ratios of the Higgs decay widths in the HSM and four types of the THDM to those in the SM, defined in Eq.~\eqref{delR}, between $h\to\tau\tau$ and $h\to ZZ^*$. 
 See Fig.~\ref{fig:dR_tb} for the descriptions.}
\label{fig:dR_tz}
\end{figure}

In Fig.~\ref{fig:dR_tz} we show the correlation in radiative corrections to the decay rates of $h\to\tau\tau$ and $h\to ZZ^*$, 
where the descriptions for the figure are same as in Fig.~\ref{fig:dR_tb}.
Similar to the discussion on Fig.~\ref{fig:dR_tb},
patterns of the deviations can be mainly governed by the mixing effects on the Higgs couplings at the tree level.
The HSM can be distinguished from the THDMs if the deviation $\Delta R_{ZZ^*}$ is larger than a few per cent.
On the other hand, the results in the Type-I (Type-II) THDM coincide with those 
in the Type-Y (Type-X) THDM at the tree level, as the Yukawa structures for leptons are the same.
An important remark on the THDMs is the difference of the magnitude of the deviations between $h\to ff$ and $h\to VV^*$~\cite{Kanemura:2015mxa}.
The 5\% deviation for the coupling in the gauge sector (i.e. $s_{\beta-\alpha}=0.95$) gives rise to $\Delta R_{ZZ^*} \sim -10\%$ -- $-15\%$, while $\Delta R_{ff}\sim \pm60\%$ or more. 
We also remark that the dependence on the extra Higgs boson masses are different between the HSM and the THDMs.
The deviations in the HSM can be larger than those in the THDMs for $m_H>1$~TeV~\cite{Kanemura:2015fra,Kanemura:2017wtm}.

Comparing with the $h\to ff$ decay, the loop correction to the three-body $h\to ZZ^*\to Zff$ decay is much more intricate as discussed in Sec.~\ref{sec:hvv}.
The pure quantum effects from the extra Higgs bosons can be seen as thickness of the line in the HSM. 
In the THDMs, the pure loop effects can be in the regions
below the tree-level $\tb=3$ line in the Type-I THDM and 
below the $\tb=1$ line in the Type-II THDM.
As seen, the loop corrections from the heavy Higgs bosons tend to reduce the decay widths.
We find that in the THDMs the corrections to $\Delta R_{ZZ^*}$ are larger for a smaller $\tb$ and for a smaller $\sba$.
We find that 
the magnitude of the radiative corrections can be as large as a few times per cent. 
In this plot, near the origin, i.e. around the SM-like limit,
we clearly observe the non-decoupling effect discussed above. 
In contrast to the tree-level analysis, 
even with heavier Higgs bosons a larger deviation can be predicted  
as seen in the region with $m_H=700$~GeV near the SM-like limit in the THDMs. 

\begin{figure}
 \center 
 \includegraphics[width=0.49\textwidth]{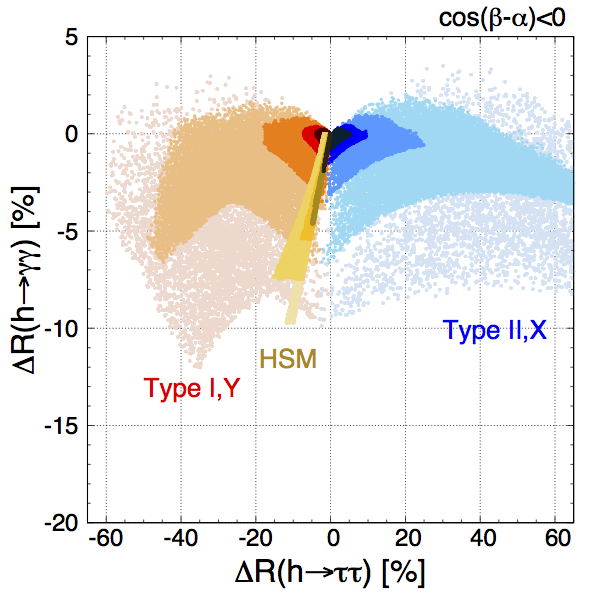}\ \
 \includegraphics[width=0.49\textwidth]{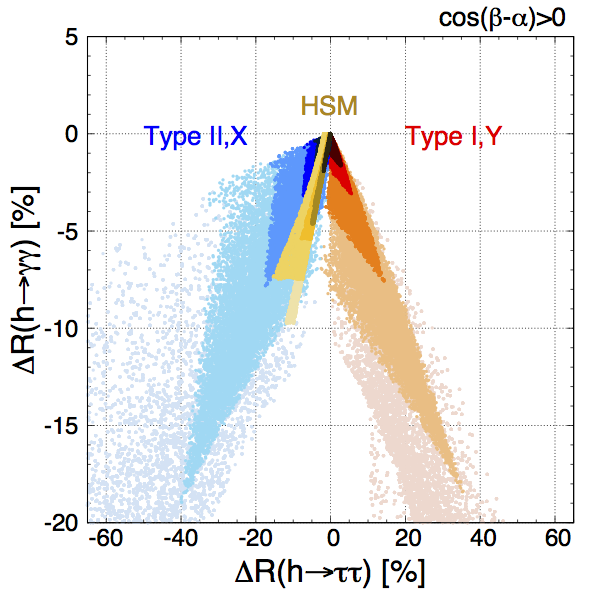} 
 \caption{Correlation of the ratios of the Higgs decay widths in the HSM and four types of the THDM to those in the SM, defined in Eq.~\eqref{delR}, between $h\to\tau\tau$ and $h\to\gamma\gamma$. 
 See Fig.~\ref{fig:dR_tb} for the descriptions.}
\label{fig:dR_tg}
\end{figure}

\subsection{$h\to\tau\tau$ vs. $h\to\gamma\gamma$}

In Fig.~\ref{fig:dR_tz} we show correlations between $h\to\tau\tau$ and $h\to\gamma\gamma$.
Unlike the previous correlations, the shape of the regions predicted in the THDMs is rather different between the cases for different sign of $c_{\beta-\alpha}$.
The reason why the deviations are larger for $\cba>0$ than those for $\cba<0$ is that contributions of
the charged Higgs loop diagram and the top-loop diagram are constructive for $\cba>0$ so that the deviations from the SM prediction can be larger. 
The different shapes of the predicted regions between $c_{\beta-\alpha}<0$ and $>0$ can help to disentangle the degeneracy between the Type-I THDM and the Type-II THDM. 
For instance, 
$\Delta R_{\gamma\gamma}$ is mostly negative, but can be positive by a few per cent at most for $c_{\beta-\alpha}<0$.
In such a case, if $\Delta R_{\tau\tau}$ is negative (positive), the model can be identified as 
the Type-I (Type-II) THDM.

\section{Conclusions}\label{sec:summary}

In order to examine the Higgs sector using future precision data, 
we have studied the deviations from the SM predictions for
the various partial decay widths of the 125~GeV Higgs boson 
with one-loop EW and one-loop QCD corrections 
in extended Higgs models, such as
the HSM as a real singlet extension and four types of the THDM with softly broken $Z_2$ symmetry.

By employing and extending \hcoup\!\!, which evaluates the renormalized vertex functions for the discovered Higgs boson in various extended Higgs models, 
we calculated the partial decay widths for $h\to\tau\tau$, $bb$, $cc$, $Zff$ and $\gamma\gamma$, 
and presented the ratios of the decay widths in the extended Higgs models to those in the SM.

We described the pattern of the deviations in the various correlations, which 
are distinctive for each model and determined by the tree-level mixing effects on the Higgs couplings. 
Even with a full set of radiative corrections 
we may be able to discriminate these extended Higgs models
as long as any of the deviations from the SM predictions is detected at future precision measurements.
For example, we can discriminate the Type-X THDM and the Type-Y THDM from the others (the Type-I THDM, the Type-II THDM and the HSM) by the analysis shown in Fig.~\ref{fig:dR_tb}.
Then, the Type-II THDM can be separated from the Type-I THDM and the HSM as shown in Fig.~\ref{fig:dR_tc}.
We can further distinguish the Type-I THDM from the HSM as in Fig.~\ref{fig:dR_tz}.
Finally, by using the data from $\Gamma(h\to\gamma\gamma)$, we can check the consistency as in Fig.~\ref{fig:dR_tg}. 
In order to complete such a logic realized, we need to measure the decay rate of $h\to cc$ precisely, in addition to measuring the other decay modes as precisely as possible.  
To this end the realization of future lepton colliders is deeply desirable.
Furthermore, although the dependence on the additional Higgs boson mass generally indicates the decoupling behavior under the theoretical constraints from perturbative unitary and vacuum stability, the loop effects from the extended Higgs sector can be large due to the non-decoupling effect.
We can extract important information on the mass scale of extra Higgs bosons indirectly from
the magnitude of the deviations of the SM-like Higgs decay widths,
even if new particles are not discovered at the LHC.

\begin{center}
{\bf Acknowledgments}
\end{center}
%
The work of S.K. was supported, in part,
by Grant-in-Aid for Scientific Research on Innovative Areas, 
the Ministry of Education, Culture, Sports, Science and Technology, No.~16H06492,
Grant H2020-MSCA-RISE-2014 No.~645722 (Non-Minimal Higgs), 
and JSPS Joint Research Projects (Collaboration, Open Partnership)
``New Frontier of neutrino mass generation mechanisms via Higgs physics at LHC and flavour physics''. 


\providecommand{\href}[2]{#2}\begingroup\raggedright\endgroup

\end{document}